\documentstyle[preprint,aps,eqsecnum]{revtex}

\newcommand\ignore[1]{}

\begin{document}

\draft
\title{Vortex lattice transition in $D$-wave superconductors}
\author{Jun'ichi Shiraishi${}^1$,
Mahito Kohmoto${}^1$ and Kazumi Maki${}^2$}
\address{${}^1$Institute for Solid State Physics,
University of Tokyo, Roppongi, Minato-ku, Tokyo 106, Japan}
\address{${}^2$Department of Physics and Astronomy,
University of Southern Calfornia Los Angeles,
Cal. 90089-0484, USA}

\maketitle
\begin{abstract}
Making use of the extended Ginzburg Landau theory, which includes the
fourth order derivative term, we study the vortex state in a magnetic
field parallel to the $c$ axis. The vortex core
structure is distorted due to this higher order
term, which reveals the fourfold symmetry.
Further, this distortion gives rise to the core interaction
energy, which favors a square lattice tilted by $45^\circ$ from the
$a$ axis.
The critical field of this transition is determined.
The magnetization
diverges at the transition. This suggests the
transition is of the first order.
\end{abstract}
\pacs{ }

\narrowtext

\section{Introduction}
After a few years of controversy, $d$-wave superconductivity in the
hole-doped high $T_c$ cuprates appears to be finally
established\cite{GEN,MW}.
However, the electron-doped high $T_c$ cuprates appear to be described by $s
$-wave superconductivity\cite{A,S}.

$D$-wave superconductivity manifests itself
as fourfold symmetry of the  vortex state
when a magnetic field is applied either
parallel to the $c$ axis or within the $a$-$b$ plane\cite{MSW}.
In particular to the study of the vortex lattice in the vicinity of
the
upper critical field\cite{WM} and the quasi-particle spectrum around a
single vortex\cite{MScW,I} in a magnetic field
parallel to the $c$ axis indicate that the square vortex
lattice tilted by $45^\circ$ from the
$a$ axis should be most stable except in the immediate vicinity
of the superconducting transition temperature $T_c$.
Indeed such a square lattice,
though elongate in the  $a$ direction has been
seen in YBCO monocrystals by small angle neutron scattering
(SANS)\cite{K} and scanning tunneling microscopy (STM)\cite{M}
at low temperature and in a low magnetic field.
On the other hand, the fourfold symmetry predicted
for the density of states near the vortex core appeared
not to have been seen by STM\cite{M} in YBCO monocrystals.
This, we believe, indicates the failure of the quasi-classical
approximation used in these theoretical analysis.
Indeed, recent studies\cite{MKM,Y} of the Bogoliubov-de Gennes (BdG)
equation clearly indicate not only the breakdown
of the quasi-classical approximation for YBCO, but also the presence of the
extended states with small energies (say $|E|<0.1 \Delta$)
which exhibits clearly the fourfold symmetry anticipated from
the square vortex lattice.

More recently a very similar square vortex lattice have been seen in
ErNi${}_2$B${}_2$C, YNi${}_2$B${}_2$C and LuNi${}_2$B${}_2$C
by SANS\cite{Y,Ye} and in YuNi${}_2$B${}_2$C by STM imaging\cite{D}.
Although superconductivity in borocarbides is believed to be
conventional $s$-wave\cite{E}, the above square lattice
together with the presence of antiferromagnetic
phase in closely related borocarbides suggest strongly that
superconductivity in
borocarbides will be of $d$-wave as well\cite{NITC}.
Incidentally the square
vortex lattice and related vortex lattice transition have
recently been studied using the generalized London equation\cite{D,Af,Ko}.
The phenomenological free energy used by these authors
resembles the one for $d$-wave superconductivity.

The object of this paper is twofold. (i) Making use of
the extended Ginzburg-Landau equation, we first study a single
vortex line in a magnetic field parallel to the $c$ axis.
Unlike Refs. \cite{Af,Ko} we believe that the modification of
the
vortex core structure is of prime importance. Indeed, the vortex
exhibits the fourfold symmetry, which will
have a number of consequences. For example
it will modify the quasi-particle spectrum around a vortex.
One more significant fact is that
this will generate vortex core interaction
energy, which favors the alignment of two vortices either parallel
to $(1,1,0)$ or $(1,-1,0)$. Indeed a similar vortex
solution has been found numerically previously by
Enomoto {\it et. al.}\cite{En}.
But our analytical result is of prime importance in the following.
(ii) From a study of the two-vortex problem, we consider
the vortex lattice for a class of isoceles.
We find in the low field limit ({\it i.e.} $B\simeq H_{c1}(t)$)
the vortices form a triangular lattice as in a conventional $s$-wave
superconductor. When the magnetic field increases, the triangular
lattice transforms first gradually and then suddenly
to the square lattice when $B=H_{cr}$.
In the temperature range not very far from $T_c$
({\it i.e.} ${1\over 2}T_c<T<T_c$) we predict
\begin{equation}
H_{cr}=0.524 (-\ln t)^{-1/2} \kappa^{-1} H_{c2}(t),
\end{equation}
where $t=T/T_c$ and $\kappa$ is the Ginzburg-Landau
parameter. Though the $B$-dependence of the
apex angle $\theta$ we obtained is rather similar to the
ones obtained in Refs.\cite{Af,Ko}, the detail is quite
different.
For example we find the vortex lattice transition is of the first order in a
sharp
contrast to Ref. \cite{Af,Ko}, where it is found of the second order. Furthe
r the present
model describes $\theta$ dependence on $B$
more consistent with SANS result\cite{Y} than that of Ref.\cite{Ko},
which may suggest that the core interaction between two vortices will
be more critical than the term arising from the anisotropy of the
magnetic interaction considered in Refs.\cite{Af,Ko}.
Unfortunately the related SANS study for high $T_c$ cuprates is not
available at this time of writing.

\newcommand{\br}{{\bf  r}}
\section{Extended Ginzburg-Landau equation and single vortex problem}
We consider a weak-coupling model for $d$-wave superconductors\cite{WM2}.
Extending the procedure used by Ren {\it et.al.}\cite{R}, we obtain;
\begin{eqnarray}
&&\Biggl( -\ln t  + {7\zeta(3) \over 2 (4 \pi T)^2} v^2
 (\partial_x^2+\partial_y^2)
 + {31\zeta(5) \over 16 (4 \pi T)^4} v^4
\left[
5(\partial_x^2+\partial_y^2)^2\right.\nonumber\\
&&\qquad\qquad\left.+2(\partial_x^2-\partial_y^2)^2 \right]
\Biggr) \Delta(\br)={21\zeta(3) \over  (4 \pi T)^2}
|\Delta(\br)|^2 \Delta(\br), \label{GL}
\end{eqnarray}
which is converted into the dimensionless form
\begin{eqnarray}
&&\left( 1 +
 (\partial_x^2+\partial_y^2)
+ \epsilon
\left[
5(\partial_x^2+\partial_y^2)^2+2(\partial_x^2-\partial_y^2)^2 \right]
\right) \Delta(\br) =
|\Delta(\br)|^2 \Delta(\br). \label{GL0}
\end{eqnarray}
where we have introduced
$$
\xi(T)^2= { 7 \zeta(3)v^2\over 2 (4 \pi T)^2 (-\ln t)},\quad
\Delta(T)^2={  (4 \pi T)^2 (-\ln t)\over  21\zeta(3)},
$$
$t=T/T_c$, and
rescaled $\br\rightarrow \xi(T) \br $,
$\Delta(\br)\rightarrow \Delta(T) \Delta(\br)$.
Here $\partial_x$ and $\partial_y$ are gauge invariant differential
operators and we define the  small parameter
$
\epsilon\equiv {31\zeta(5)(-\ln t)/ 196 \zeta(3)^2} \sim 0.114(-\ln t).
$

Equation (\ref{GL}) is written down basically in \cite{En},
though we ignore a few terms of the order of $(-\ln t)^2$
since they are of secondary importance in what follows.
Here we concentrate on the effect of the $\epsilon$-term, which is the basic
symmetry breaking term.

Assume that $\Delta(\br)$ is given by
\begin{eqnarray}
\Delta(\br) &=&
g(r)e^{i \phi}
+ \epsilon \left(
e^{4 i \phi} \alpha(r)+e^{-4 i \phi}\beta(r)+\gamma(r)\right) e^{i
  \phi}.
\end{eqnarray}
Substituting this in Eq. (\ref{GL0}) we find $g(r)$ for
$r \gg 1$;
\begin{eqnarray}
g(r)&=&1
-{1\over 2} r^{-2}
-{9\over 8} r^{-4}
-{161\over 16} r^{-6} \cdots,
\end{eqnarray}
and equations for $\alpha(r),\beta(r)$ and $\gamma(r)$
for
$r \gg 1$;
\begin{eqnarray}
A(r)+
\left( 1 +
\left(\partial_r^2+{1 \over r} \partial_r -{25\over r^2}\right)
\right)\alpha(r)&=&g(r)^2 (2\alpha(r) +\beta(r)),\label{alphaeq}\\
B(r) +
\left( 1 +
\left(\partial_r^2+{1 \over r} \partial_r -{9\over r^2}\right)
\right)\beta(r)
&=&g(r)^2
(\alpha(r)+ 2\beta(r)),\label{betaeq}\\
C(r) +
\left( 1 +
\left(\partial_r^2+{1 \over r} \partial_r -{1\over r^2}\right)
\right)\gamma(r)
&=& g(r)^2 3 \gamma(r),\label{gamma}
\end{eqnarray}
where
\begin{eqnarray}
A(r)&=& {105 \over 2} r^{-4}-{945 \over 4} r^{-6}
-{31185 \over 16}r^{-8}-
{1450449\over 32}r^{-10}\cdots,\label{A}\\
B(r)&=& -{15 \over 2} r^{-4}-{105 \over 4} r^{-6}
-{8505 \over 16}r^{-8}-
{557865 \over 32}r^{-10}\cdots,\label{B}\\
C(r)&=&  -18 r^{-4}-135 r^{-6} -{14175 \over 4}r^{-8}-
{1065015\over 8}r^{-10}\cdots.\label{C}
\end{eqnarray}
Then we find
\begin{eqnarray}
\alpha(r)&=& {5 \over 2} r^{-2} +
\left(c-{55 \over 4} \log r \right) \; r^{-4}
+
\left({-2873-456 \;c \over 80}+{627 \over 8} \log r \right) \; r^{-6}
\cdots, \label{alphainf}\\
\beta(r)&=& - {5 \over 2} r^{-2} +
\left({5 -2\;c\over 2}+{55 \over 4} \log r  \right) r^{-4}
+
\left({-6627-184 \;c \over 80}+{253 \over 8} \log r \right) \;
  r^{-6}\cdots,\label{betainf}
\end{eqnarray}
and
\begin{eqnarray}
\gamma(r)&=& - 9 r^{-4}-{ 297 \over 2}r^{-6}-{5313\over 8}r^{-8}
\cdots.
\end{eqnarray}
In this solution we find a free parameter $c$, which fortunately
does not show up in the core interaction term which we are going to
discuss in the following section.
Note that the choice $c=5/4$ makes the first few terms symmetric;
$ \alpha(r)= 5/2 r^{-2}+(1-11 \log r) 5/4 r^{-4}\cdots,
  \beta(r)= -5/2 r^{-2}+(1+11 \log r) 5/4 r^{-4}\cdots$.
We will also discuss in the next paragraph that the choice $c\sim 5/4$
is necessary to have approximate solutions.

For later purposes, it is convenient to introduce the
interpolation expressions which give the correct asymptotics for
 $r \rightarrow 0$. We find
\begin{eqnarray}
g(r)
&=&\tanh {r\over c_0}
-{1\over 2 r^2} \left(1-c_1{\rm sech}{r\over c_0}\right)
\tanh^5 {r\over c_0}
-{9\over 8 r^4} \left(1-c_2{\rm sech}{r\over c_0}\right)
\tanh^9 {r\over c_0}\cdots,
\end{eqnarray}
\begin{eqnarray}
\alpha(r)&=& {5 \over 2} r^{-2} \tanh^7 {r \over c_3}+
\left({5 \over 4}-{55 \over 4} \log r \right) \; r^{-4}
\tanh^{11} {r \over c_3}\cdots,\label{alpha}\\
\beta(r)&=& - {5 \over 2} r^{-2} \tanh^5 {r \over c_3}+
\left({5 \over 4}+{55 \over 4} \log r  \right) r^{-4}
\tanh^9 {r \over c_3}\cdots, \label{beta}
\end{eqnarray}
where $c_0=1.71$, $c_1=0.80$, $c_2=1.35$.
The way to fix these constants is the
following. Using the GL equation (\ref{GL0}), we can express
all the constants $c_1,c_2\cdots$ by $c_1$.
The constant $c_1$ can be obtained by performing numerical
integration of the GL equation with the boundary conditions
$g(0)=0,\lim_{r\rightarrow \infty} g(r)=1$. In principle,
we can apply the same procedure to $\alpha(r)$ and $\beta(r)$.
However, we simply start from the ansatz (\ref{alpha}) and (\ref{beta})
which are given from (\ref{alphainf}) and (\ref{betainf})
by introducing suitable powers of ${\rm tanh}r/c_3$,
and observe that these
with $c_3 =2.5$ and $c=5/4$ agree very nicely with the
numerical results obtained by Enomoto {\it et. al.} \cite{En}.
We show in Fig.1 $\alpha(r)$ and $\beta(r)$ as function of $r$.
These are compared with
$8 f^{(1)}_1(r)$ and
$8 f^{(1)}_{-1}(r)$ in Enomoto {\it et. al.}.
We see our analytic expressions are very close to the
numerical ones in Enomoto {\it
  et. al}. We have not shown $\gamma(r)$ as this term is somewhat
different
from the one in Enomoto {\it et. al.} since our starting equation is
different.

\section{Interaction between two vortices}
Before studying the regular vortex lattice, let us
consider the two-vortex problem.
We assume that two vortices are
 placed at
$(0,0)$ and $(d \cos \theta,d\sin \theta)$ and ($\kappa\gg d\gg 1$).
The free energy in dimensionless units is given by
\begin{eqnarray}
\Omega&=&
\int d^2 r\;
\left( -|\Delta|^2 +|\partial_x \Delta|^2+|\partial_y
  \Delta|^2\right.\nonumber\\
&&\left. -
\epsilon |(5(\partial_x^2+\partial_y^2)+2(\partial_x^2-\partial_y^2) )
\Delta|^2+{1 \over 2} |\Delta|^4+{1 \over 8\pi} b^2
\right)\\
&=&
\int d^2 r\;
\left( -{1 \over 2} |\Delta|^4+{1 \over 8\pi} b^2
\right), \nonumber
\end{eqnarray}
where $b=b(\br)$ is the local magnetic field.
Making use of the usual approximation
$\Delta({\bf r})= \Delta \prod_i f(\br-{\bf r}_i)$, where
\begin{eqnarray}
f(\br)=\left(  g(r)+\epsilon(\alpha(r)e^{4 i\phi}+\beta(r)e^{-4
  i\phi}+
\gamma(r))
\right)e^{i\phi},
\end{eqnarray}
is the single vortex solution, $g(r)\sim {\rm tanh}\, r$
and neglecting $\gamma(r)$ which is irrelevant for the
fourfold symmetry, we obtain
\begin{eqnarray}
\Omega_{\mbox{ two-vortex}}&\simeq&-{1 \over 2} \int d^2r \;
(\tanh r+ \epsilon \cos 4\phi (\alpha(r)+\beta(r)))^4\nonumber\\
&&\qquad\quad\times
(\tanh r'+ \epsilon \cos 4\phi' (\alpha(r')+\beta(r')))^4 \nonumber\\
&\simeq&
-{1\over 2}\left(A- 2 a_1 -2 a_1\epsilon (\alpha(d)+\beta(d))
\cos 4\theta
  \right),
\end{eqnarray}
where $A$ is the area and
\begin{eqnarray}
a_1= \int d^2r \; \left(2 {\rm sech}^2 r -{\rm sech}^4 r \right)
={8\pi \over 3} \left(\ln 2 + {1 \over 8}\right)\simeq 6.854.
\end{eqnarray}

On the other hand the magnetic interaction between two vortices is
given by
${2\pi \over \kappa^2}K_0({d \over \kappa})$ (the
London formula) where $K_0(z)$ is the modified Bessel
function. Strictly speaking the magnetic interaction is also modified due to
the higher order term (see, for example, Ref. 18). Indeed the correction
term decays like $d^{-2}$ with $d$, but this term does not contain the extra
$\kappa$-dependence. Therefore the correction term to the magnetic
interaction is completely negligible when $\kappa \gg 1$ as in high $T_c$ cu
prates.
Therefore the core interaction give a strongly directional energy
$\sim d^{-4}\cos 4\theta$, while the magnetic energy is isotropic as in
conventional $s$-wave superconductor.

\section{Vortex Lattice}
Let us consider a vortex lattice where lattice points are given by
$\br_{l,m}=
r_{l,m}(\cos \theta_{l,m},\sin \theta_{l,m})=
l d(\cos \theta,\sin \theta)+m d(\cos \theta,-\sin \theta)$,
where $l,m$ are integers
$
d=\sqrt{ \phi_0 \over \sin (2 \theta) B},
$
and $\phi_0$ is the flux quantum.
For later convenience, we separate the lattice into
even and odd lattices as
$\br^{(e)}_{l,m}=
r^{(e)}_{l,m}(\cos \theta^{(e)}_{l,m},\sin \theta^{(e)}_{l,m})=
2l d(\cos \theta,\sin \theta)+2m d(\cos \theta,-\sin \theta)$ and
$\br^{(o)}_{l,m}=
r^{(o)}_{l,m}(\cos \theta^{(o)}_{l,m},\sin \theta^{(o)}_{l,m})=
(2l+1) d(\cos \theta,\sin \theta)+(2m+1) d(\cos \theta,-\sin \theta)$.
Note $l$ and $m$ run over all possible {\it integers}.
Then the free energy of the vortex lattice is given by
\begin{eqnarray}
\Omega&=&-{1 \over 2}\left(  A- a_1 \xi^2 n_\phi -
\epsilon 10 a_1\xi^2 n_\phi
\sum_{l,m} {}' {\xi^4 \over r_{l,m}^4} \cos 4\theta_{l,m} \right)+
{2\pi\over \kappa^2} n_\phi \xi^2
\sum_{l,m}{}' K_0\left({r_{l,m}\over \lambda}\right),\label{omega}
\end{eqnarray}
where $n_\phi=B/\phi_0$ the vortex density per unit area.
Here we consider only the vortex core interaction between two vortices,
since the three vortex interaction is exponentially small when $d/\xi\gg 1$.
Further, we have neglected the fourfold symmetric term in the
magnetic interaction term since it is proportional
to $\epsilon/\kappa^2$.
So except for the condensation energy
$(-{1 \over 2} A)$, the second term and the last term are
proportional to $B$, while the core interaction  energy (the third
term)
is proportional to $B^3$.
As the magnetic field increases from $B=H_{c1}(t)$,
the third term becomes more and more dominant and for
$B\geq H_{cr}$ the square vortex lattice will be established.
The last term in Eq. (\ref{omega}) contains the sum
\begin{eqnarray*}
&&\sum_{l,m\in{\bf Z}\atop p=e,o}\!\!{}'
K_0\left({r^{(p)}_{l,m}\over \lambda}\right)
=\sum_{l,m}{}' K_0\left((l^2 \mu^2+ m^2{\mu'}^2)^{1/2}\right)+
\sum_{l,m} K_0\left(((l-1/2)^2 \mu^2+ (m-1/2)^2{\mu'}^2)^{1/2}\right),
\end{eqnarray*}
where $\mu=2d\sin \theta/\lambda,\mu'=2d\cos \theta/\lambda$.
Following the argument by Fetter {et.al.}\cite{Fet},
namely, using the integral representation of the function $K_0(x)$
and two Poisson summation formulas (see Appendix),
we can rewrite these infinite summations.
Then the last term in Eq. (\ref{omega})
becomes (for $\lambda \gg d$)
\begin{eqnarray}
&&{2\pi\over \kappa^2} n_\phi\xi^2 \sum_{l,m}{}'K_0\left(r_{l,m} \over
\lambda\right) \nonumber \\
&\simeq &
{2\pi\over \kappa^2} n_\phi\xi^2 \left[
{4\pi \over \mu\mu'}+
{1 \over 2} \ln {\mu\mu'\over 4\pi}-{1 \over 2}(1-\gamma) \right.\\
&&\left.+{1\over 2} \sum_{l,m}{}'
\left( E_1\left(\pi(l^2{\mu \over \mu'}+m^2{\mu' \over \mu})\right)
+{(-1)^{l+m}+
 \exp \left(-\pi (l^2{\mu' \over \mu}+m^2{\mu \over \mu'})   \right)
  \over
\pi (l^2{\mu' \over \mu}+m^2{\mu \over \mu'})}
 \right)
\right].\nonumber
\end{eqnarray}
The angle $\theta_{\rm min}$ which minimizes the free energy is
obtained by studying the function
\begin{eqnarray*}
f(\theta)&=&\left({B\over H^*(t)}\right)^2\sum_{l,m}{}'
{\sin^2 2 \theta   \cos 4 \theta_{l,m} \over
\left((l+m)^2 \sin^2 \theta+(l-m)^2 \cos^2 \theta\right)^2}\\
&&+ \sum_{l,m}{}'
\left( E_1\left(\pi(l^2{\tan\theta }+
m^2{\cot \theta})\right)
+{(-1)^{l+m}+
 \exp \left(-\pi (l^2{\cot \theta}+m^2{\tan\theta})   \right)
  \over
\pi (l^2{\cot\theta}+m^2{\tan\theta})}
 \right),
\end{eqnarray*}
where
\begin{eqnarray*}
H^*(t)= \left({98 \zeta(3)^2 (2\pi)^3 \over
155 a_1 \zeta(5) (-\ln t) }\right)^{1/2} {H_{c2}(t) \over \kappa}
\sim 5.64667 (-\ln t)^{-1/2} {H_{c2}(t) \over \kappa}.
\end{eqnarray*}
Then the minimization of $f(\theta)$ gives
Fig. 2 where the apex angle $\theta_{\rm min}$ is shown as function of
$B/H_{cr}$ where
\begin{equation}
H_{cr}=0.524 (-\ln t)^{-1/2} \kappa^{-1} H_{c2}(t).
\end{equation}
For $B \geq H_{cr}$ the square lattice is fully established. Note
also that ${d\theta / dB}$ diverges at $B=H_{cr}$
indicating the possible phase transition.
Earlier a similar $\theta$-$B$ curve was obtained within
the generalized London equation\cite{Af,Ko}.
However, the present result appears to be more consistent with
the observed $B$ dependence of $\theta$ by SANS
from ErNi${}_2$B${}_2$C at $T= 3.5 $K\cite{Y}.
Inserting $\theta$ determined thus into Eq. (\ref{omega}),
we find the free energy
\begin{eqnarray}
\Omega&=&\Omega_0+{2 \pi \xi^2 H_{cr}\over \kappa^2 \phi_0}
\overline{f}\left(B\over H_{cr}\right),
\end{eqnarray}
where the first term
\begin{eqnarray}
\Omega_0&=&
-{A\over 2} + {a_1 \xi^2 \over 2\phi_0} B+
{2\pi \xi^2 \over \kappa^2 \phi_0}B
\left[
{2\pi \lambda^2\over \phi_0} B+{1 \over 2}\log
{\phi_0\over 2\pi \lambda^2 B}-{1 \over 2} (1-\gamma)
  \right],
\end{eqnarray}
depends on $B$ in a non-singular way, and the second term is
\begin{eqnarray}
\overline{f}\left(B\over H_{cr}\right)&=&
{B\over H_{cr}}
f\left(\theta_{\rm min}\left({B\over H_{cr}}\right)\right).
\end{eqnarray}
We show $\overline{f}(B/H_{cr})$ as a function of $B/H_{cr}$
for $0 \leq B/H_{cr}\leq 1.2$ in Fig. 3.
A cusp is observed at $B= H_{cr}$.
The magnetization $-M=\partial \Omega/\partial B$ has a
singularity at $B=H_{cr}$ due to the cusp in
$\overline{f}\left(B/ H_{cr}\right)$.
Fig. 4 shows the singular part of the
magnetization $-M_{\rm singl}\equiv\partial
 \overline{f}(B/H_{cr})/\partial (B/H_{cr})$
for $0 \leq B/H_{cr}\leq 1.2$.
Fig.3 and Fig.4 show clearly this phase transition is of the first order
with
negative latent heat ({\it i.e.} less entropy in the square lattice),
which should be readily accessible experimentally.
Suppose we choose $\theta-90^\circ$ as an order parameter,
which does not have discontinuity.
The approach to zero, however, is very sharp and it can not
be described by power law as the usual second order phase transitions.
The transition can be classified as a special kind of the first order transi
tion,
even though
the order parameter is continuous.
A detailed study on the nature of this phase transition
is under consideration\cite{MSK}.

\section{Concluding Remarks}
By analyzing the extended Ginzburg-Landau equation for $d$-wave
superconductor, we discover that
the vortex core contains a long range fourfold term
which is proportional to
$r^{-4} \cos 4 \phi$ when $r\geq \xi$.
The effect of this term on the quasi-particle
spectrum is under current study. This fourfold term
gives rise to the vortex core interaction, which favors the
orientation of two vortices parallel to the diagonal directions
$(1,1,0)$ and $(1,-1,0)$. In the low field regime we find that the
vortex lattice transforms from triangular to square
as $B$ increases and that the last transition to the
square lattices is very steep and of the first order. The present result app
ears to
describe very well the vortex transition observed in
ErNi${}_2$B${}_2$C, though the superconductivity in borocarbides is
believed to be $s$-wave.
Turning to high $T_c$ cuprates there is no similar
measurement available
even for YBCO monocrystals. On the other hand, if we
put $\kappa=100$, $H_{c2}(0)=120$ Tesla for YBCO,
we estimate $H_{cr}=1$ Tesla, which is consistent with the
observation of the square lattice at low temperature
and
in a magnetic field of a few Tesla.
Clearly a parallel
measurement of the
$B$-dependence of the apex angle $\theta$ in high $T_c$
cuprates is highly desirable.

Coming back to the
vortex lattice transformation in the vicinity of
$B\simeq H_{c2}(t)$, it is shown that  the transition is
again
continuous in contrast to an earlier analysis\cite{WM3}.
In particular the full transition to the square lattice
is completed at $t=0.81$.
Therefore it is
now possible to draw a vortex lattice phase diagram in the
$T$-$B$ plane.

We expect also that the directional core potential
not only modifies the equilibrium vortex lattice configuration
but also the collective mode,
the elastic and dynamic response of the vortex lattice.
At this moment we can say only
that $d$-wave superconductivity should bring a profound
change in our understanding of the vortex motion.

Acknowledgments\\
One of us (KM) wants to thank Japan Society of Promotion of Science
for timely support from
short term research fellowship, which enables him to spend a few weeks
at ISSP, U.of Tokyo. Also he thanks Prof. Peter Wyder and
MPI-CNRS
at Grenoble for a warm hospitality where a part of his work is done.
The present work is in part
supported by National Science Foundation under grant number DMR95-31720.

\section*{Appendix}
In this Appendix, we list some useful formulas
for studying the free energy $\Omega$ (\ref{omega}) of the vortex lattice with
the apex angle $\theta$.

We have to treat the lattice sums
\begin{eqnarray*}
\Xi^{(e)}(\mu,\mu')
&=&\sum_{l,m}{}' K_0\left((l^2 \mu^2+ m^2{\mu'}^2)^{1/2}\right),\\
\Xi^{(o)}(\mu,\mu')
&=&\sum_{l,m} K_0\left(((l-1/2)^2 \mu^2+ (m-1/2)^2{\mu'}^2)^{1/2}\right).
\end{eqnarray*}
The Poisson sum formulas
\begin{eqnarray*}
&&\sum_{l} \exp \left( -l^2 \mu^2/4 \tau \right)
= {\sqrt{4\pi \tau} \over \mu }
\sum_{l} \exp \left( -4 \pi^2 \tau l^2 \mu^2 \right),\\
&&\sum_{l} \exp \left( -(l-1/2)^2 \mu^2/4
\tau \right)
= {\sqrt{4\pi \tau} \over \mu }
\sum_{l}(-1)^{l}
 \exp \left( -4 \pi^2 \tau l^2 \mu^2 \right),\\
\end{eqnarray*}
can be obtained from Jacobi's imaginary transformations
for the elliptic theta functions;
$\vartheta_3(v,\tau)=e^{\pi i/4} \tau^{-1/2} e^{-\pi i v^2/\tau}
 \vartheta_3(v/\tau,-1/\tau)$ and
$\vartheta_4(v,\tau)=e^{\pi i/4} \tau^{-1/2} e^{-\pi i v^2/\tau}
 \vartheta_2(v/\tau,-1/\tau)$.
Using the argument by Fetter {et.al.}\cite{Fet},
we obtain
\begin{eqnarray*}
\Xi^{(e)}(\mu,\mu')
&=&{2\pi \over \mu\mu'}+
{1 \over 2} \ln {\mu\mu'\over 4\pi}-{1 \over 2}(1-\gamma)\\
&&+{1\over 2} \sum_{l,m}{}'
\left( E_1\left(\pi(l^2{\mu \over \mu'}+m^2{\mu' \over \mu})\right)
+{\exp \left(-\pi (l^2{\mu' \over \mu}+m^2{\mu \over \mu'})   \right)
  \over
\pi (l^2{\mu' \over \mu}+m^2{\mu \over \mu'})}
 \right)\\
&&
- {2 \pi \over \mu \mu'} \sum_{l,m}{}'
{1 \over
\left(1+4\pi^2 ( {l^2\over \mu^2}+{m^2\over {\mu'}^2})\right)
\left(4\pi^2 ( {l^2\over \mu^2}+{m^2\over {\mu'}^2})\right)},\\
\Xi^{(o)}(\mu,\mu')
&=&
{2 \pi \over \mu \mu'}
+ {2 \pi \over \mu \mu'} \sum_{l,m} {}'
{(-1)^{l+m} \over
\left( 4\pi^2 ( {l^2\over \mu^2}+{m^2\over {\mu'}^2})\right)}\\
&&- {2 \pi \over \mu \mu'} \sum_{l,m} {}'
{(-1)^{l+m} \over
\left(1+4\pi^2 ( {l^2\over \mu^2}+{m^2\over {\mu'}^2})\right)
\left(4\pi^2 ( {l^2\over \mu^2}+{m^2\over {\mu'}^2})\right)}.
\end{eqnarray*}

\begin{figure}

Fig. 1
Plots of $\alpha(r)$ and $\beta(r)$.

Fig. 2
Appex angle $2\theta_{\rm min}$  as a
function of $B/H_{cr}$ where $2\theta_{\rm min}=90^\circ$ and $120^\circ$
correspond to the square lattice and the triangular lattice with
hexagonal symmetry, respectively.

Fig. 3
Singular part of the free energy
$\overline{f}$   as a function of $B/H_{cr}$.

Fig. 4
Singular part of the magnetization $-M_{singl}$
 as a function of $B/H_{cr}$.

\end{figure}

\end{document}